\documentclass[twocolumn,superscriptaddress,longbibliography,
	aps,pra,preprintnumbers]{revtex4-2}

\usepackage[normalem]{ulem}
\usepackage{graphicx}
\usepackage{bm}
\usepackage{color}
\usepackage{epstopdf}
\usepackage{amsmath}
\usepackage{amssymb}
\usepackage{epstopdf}
\usepackage{lipsum}

\usepackage{multirow}

\usepackage[urlcolor=blue,colorlinks=true,citecolor=blue,linkcolor=blue,pdfstartview={FitH},bookmarks=false]{hyperref}

\graphicspath{{fig/}{./fig/}{fig2/}{./fig2/}{.}}

\begin{document}

\title{
Shiba states in systems with density of states singularities
}

\author{Surajit Basak}
\email[e-mail: ]{surajit.basak@ifj.edu.pl}
\affiliation{\mbox{Institute of Nuclear Physics, Polish Academy of Sciences, W. E. Radzikowskiego 152, PL-31342 Krak\'{o}w, Poland}}

\author{Andrzej Ptok}
\email[e-mail: ]{aptok@mmj.pl}
\affiliation{\mbox{Institute of Nuclear Physics, Polish Academy of Sciences, W. E. Radzikowskiego 152, PL-31342 Krak\'{o}w, Poland}}

\date{\today}

\begin{abstract}
Magnetic impurities placed in the superconductor can lead to emergence of the Yu--Shiba--Rusinov bound states. 
Coupling between the impurity and the substrate depends on density of states (DOS) at the Fermi level and can be tuned by DOS singularities.
In this paper, we study the role of DOS singularities using the real space Bogoliubov--de~Gennes equations for chosen lattice models.
To uncover the role of these singularities (Dirac point, van Hove singularity, or the flat band), we study honeycomb, kagome, and Lieb lattices.
We show that the properties of the Shiba state strongly depends on the type of lattice.
Nevertheless some behaviors are generic, e.g. dependence of the critical magnetic coupling on the DOS at the Fermi level.
However, the Shiba states realized in the Lieb lattice exhibit extraordinary properties, which can be explained by the presence of a few nonequivalent sublattices.
Depending on the location of the magnetic impurity in the chosen sublattice, the value of critical magnetic coupling $J_\text{c}$ can be reduced or enhanced when the flat band is located at the Fermi level.
In this context, we also present differences in the local DOS and coherence lengths for different sublattices in the Lieb lattice.
\end{abstract}

\maketitle

\section{Introduction}

\begin{figure}[!b]
\centering
\includegraphics[width=\linewidth]{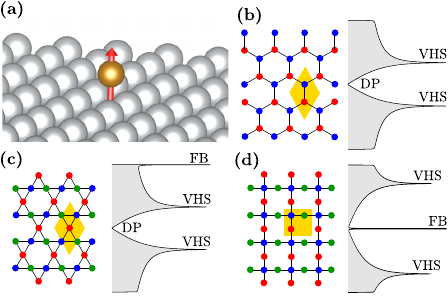}
\caption{
(a) Schematic representation of the magnetic atom (brown ball) deposited on the superconducting surface (silver balls).
Studies performed on two dimensional lattices and their density of states: (b) hexagonal, (c) kagome, and (d) Lieb lattices.
The colors of the dots (reg, green, or blue) denote non-equivalent sites, while yellow quadrangle mark the primitive unit cell.
In case of hexagonal and kagome lattice, the Dirac point (DP) is realized. 
The flat bands (FBs) are observed in the Lieb and kagome lattice.
The saddle points in the band structures give rise to van Hove singularities (VHS) in all lattices.
\label{fig.schemat}
}
\end{figure}

Interplay between superconducting system and magnetic impurity can lead to emergence of the Yu--Shiba--Rusinov~(YSR) bound states~\cite{yu.65,shiba.68,rusinov.69} (the Shiba states for short), due to local breaking of the Cooper pairs by magnetic moment of the impurity [Fig~\ref{fig.schemat}(a)].
This leads to the formation of in-gap states inside the superconducting gap, with spatially oscillating wavefunction~\cite{fetter.65}.
Recent progress in the experimental techniques has resulted in increased experimental~\cite{heinrich.pascual.18} as well as theoretical~\cite{balatsky.vekhter.06} attention in this field.

Dimensionality of the system plays a critical role in formation of the Shiba state.
The three dimensional conventional superconductor shows a fast decay of the YSR states away from the magnetic impurity.
Contrary to this, in case of two dimensional systems the YSR states are characterized by the long range coherence length~\cite{menard.guissart.15,kim.rozsa.20}

The YSR states can be observed experimentally within the topographic scanning tunneling microscopic (STM) imaging of the surface. It was first reported in presence of Mn and Gd adatoms on the surface of single-crystal Nb sample~\cite{yazdani.jones.97}.
Increased resolution allows mapping of the YSR states occurring from individual orbitals of the atom. Such cases have been reported in transition metal atoms deposited on conventional superconductors (Pb~\cite{ji.zhang.08,ruby.pientka.15,ruby.peng.16,choi.rubio-verd.17,song.park.21} or Nb~\cite{odobesko.disante.20,kuster.brinker.21,friedrich.boshuis.21,beck.schneider.21}).
YSR states were also realized by depositing Fe on NbSe$_{2}$ surface~\cite{menard.guissart.15,senkpiel.rubioverdu.19,liebhaber.acerogonzalez.20,yang.yuan.20}.
Similar observations can be made in presence of the magnetic molecules, like (Mn, Cu, V or Co) phthalocyanine~\cite{franke.schulze.11,hatter.heinrich.15,hatter.heinrich.15,etzkorn.eltschka.18,malavolti.briganti.18,brand.gozdzik.18,kezilebieke.dvorak.18,kezilebieke.zitko.19,song.park.21,song.park.21} or (Fe or Mn) porphyrin~\cite{hatter.heinrich.17,farinacci.ahmadi.20,rubioverdu.zalivar.21}.
Also, recently the fabrication of YSR states in the iron based unconventional superconductor attracted a lot of attention~\cite{wang.wiebe.21,chatzopoulos.cho.21,song.martiny.20}.

The presence of artificial structures of magnetic atoms can lead to emergence of in-gap Shiba bands.
If the non-trivial topological phase is realized~\cite{kobialka.piekarz.20}, the Majorana end modes can occur~\cite{kitaev.01}.
Recently this type of structures were realized in many experiments~\cite{kamlapure.cornils.18,kamlapure.cornils.21,ding.hu.21,liebhaber.rutten.21,schneider.beck.21,mier.hwang.21,schneider.beck.21b}.

\paragraph*{Motivation.} ---
In the simplest case, 
the magnetic impurity gives rise to a pair of in-gap states, symmetric in energy with respect to the Fermi level, characterized as YSR states.
For the classical spin, the energies of the in-gap YSR states are given as:
\begin{eqnarray}
\label{eq.eingap} E_\text{YSR} = \pm \Delta \frac{ 1 - \alpha^{2} }{ 1 + \alpha^{2} } ,
\end{eqnarray}
where $\alpha^{2} = \pi N(E_\text{F}) J$ is the dimensionless impurity coupling (neglecting Coulomb scattering), $\Delta$ is the superconducting gap, $N(E_\text{F})$ denotes the density of states (DOS) at the Fermi level $E_\text{F}$, while $J$ describes coupling between magnetic moment and electrons.
At $J = J_\text{c}$ the YSR states cross the Fermi level [$E_\text{YSR}(J_{c})=0$]. That point is related to the quantum phase transition (QPT), also known as the $0-\pi$ transition.
During the QPT, the ground state is changed from the BCS-type spinless state (for the weak coupling $J < J_\text{c}$) to the singly occupied (spinfull) configuration (for strong coupling $J > J_\text{c}$)~\cite{glodzik.ptok.18}.
As we can see from Eq.~(\ref{eq.eingap}), the value of $J_\text{c}$ is proportional to $1/N(E_\text{F})$.
Indeed, previous study of the YSR states in presence of the van Hove singularity (VHS) show that the tuning of $N(E_\text{F})$ by the VHS can lead to enhanced $J_\text{c}$~\cite{uldemolins.mesaros.21}.
Similar observation was reported in case of the critical temperature of the {\it s-wave} and {\it d-wave} superconductors on square lattice, where it was shown that $T_\text{c}$ can be tuned by increasing $N(E_\text{F})$~\cite{ptok.rodriguez.18}.
Similarly for the hexagonal lattice, in presence of the magnetic field, the VHS can lead to the superconductivity reentrant behavior~\cite{cichy.ptok.18}.

General behavior of the in-gap energies is well known from the milestone works of L. Yu, H. Shiba, and A. I. Rusinov~\cite{yu.65,shiba.68,rusinov.69}.
The general $J$-dependence of the bound state (in-gap) energies is given by Eq.~(\ref{eq.eingap}).
Similar behavior is observed also in the presence of several atoms, e.g. chain system~\cite{bjornson.balatsky.17,mohanta.kampf.18}, where in-gap states cross the Fermi level a few times.
Moreover, this type of behavior of the in-gap state energies was observed experimentally in the case of magnetic molecule of manganese phthalocyanine (MnPc) on Pb(111) surface~\cite{hatter.heinrich.15}.

Another aspect of the Shiba states is strongly associated with the Fermi surface of the studied system~\cite{ortuzar.trivini.21}.
Interestingly, the pattern of localized state induced by the magnetic impurity reflects some properties of the Fermi surface of the system~\cite{weismann.wenderoth.09}.
This is well visible in the star-shaped localization of the Shiba states around the magnetic impurity on NbSe$_{2}$~\cite{menard.guissart.15} or La~\cite{kim.rozsa.20} surface, that is associated with a six-fold symmetry of the Fermi surface of these systems.

In this paper we study the role of DOS singularities on the YSR states using exact lattice models [Fig.~\ref{fig.schemat}(b)-(d)].
These techniques allow us to study not only VHS [realized e.g. in the honeycomb lattice [Fig.~\ref{fig.schemat}(b)], but also the role of flat bands [realized e.g. in kagome or Lieb lattice~\cite{lieb.89}, presented in Fig.~\ref{fig.schemat}(c) and~\ref{fig.schemat}(d), respectively].
The flat bands can play an important role in context of the recently discovered superconducting kagome systems (like $A$V$_{3}$Sb$_{5}$~\cite{ortiz.gomes.19} or LaRu$_{3}$Si$_{2}$~\cite{mielke.qin.21}), artificial structures (like twisted bilayer graphene~\cite{cao.fatemi.18,yankowitz.chen.19}, or some type of heterostructures and interfaces~\cite{yilmaz.tong.21}).
Also, recent progress in realization of the artificial lattices~\cite{slot.gardenier.17} opens up new opportunity to study the YSR states in the flat band systems.

However, the DOS does not contain full information about the lattice, that can be important in context of correct description of the Shiba states in real systems.
As an example, this can be important in context of recent study of the Shiba states based on the Green function approach~\cite{uldemolins.mesaros.21,ortuzar.trivini.21}.
In our study we analyze this problem, based on the real space tight-binding formulation.
Our finding shows important role played by the sublattices present in the system.
Depending on the position of the magnetic impurity, the Shiba states can exhibit ``extreme'' behaviors, even within one specific lattice.

The paper is organized as follows.
Theoretical background is presented in Sec.~\ref{sec.theo}.
In Sec~\ref{sec.num} we present the numerical results and discussions.
We conclude our study in Sec.~\ref{sec.sum}.


\begin{figure*}
\centering
\includegraphics[width=\linewidth]{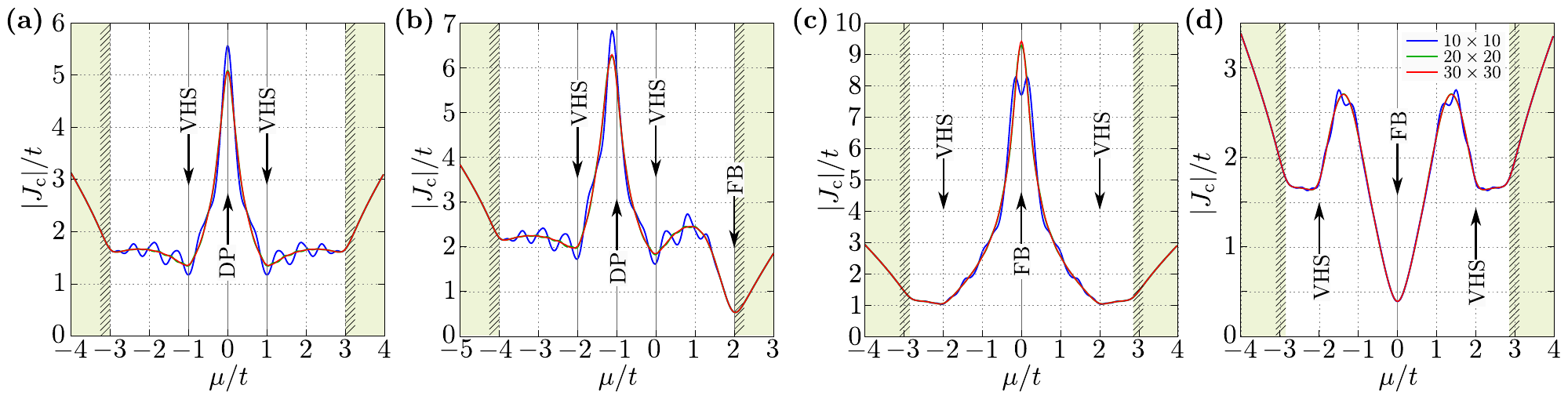}
\caption{
Critical value of the magnetic coupling $J$, denoting phase transition, for different size of lattices (as labeled).
Results are shown for: (a) hexagonal, (b) kagome, and (c,d) Lieb lattices.
For the Lieb lattice, the impurity is located at (c) the corner site or at (d) the edge site [i.e. blue and green/red site in Fig.~\ref{fig.schemat}(d), respectively].
The chemical potentials for flat bands, van Hove singularities, and Dirac points, are marked as FB, VHS, and DP, respectively.
Ranges of chemical potential which are out-of-range of the bands for given lattices are marked by green areas.
\label{fig.jc}
}
\end{figure*}

\section{Theoretical background}
\label{sec.theo}

The system is described by the Hamiltonian:
\begin{eqnarray}
\label{eq.ham}
H = H_{0} + H_\text{SC} + H_\text{imp} .
\end{eqnarray}
The first term describes the tight binding model of the lattice (cf. Fig.~\ref{fig.schemat}):
\begin{eqnarray}
H_{0} = -t \sum_{\langle ij \rangle \sigma} \hat{c}^{\dagger}_{i \sigma} \hat{c}_{j \sigma} - \mu \sum_{i \sigma} \hat{c}^{\dagger}_{i \sigma} \hat{c}_{i \sigma}
\end{eqnarray}
where $\hat{c}^{\dagger}_{i \sigma}$ ($\hat{c}_{i \sigma}$) denotes the creation (annihilation) operator an electron with spin $\sigma$ at site $i$,
$t$ is the hopping integral between nearest neighbors $\langle i,j \rangle$, and $\mu$ is the chemical potential. 
The second term is responsible for superconductivity:
\begin{eqnarray}
H_\text{SC} = \sum_{i} \left( \Delta \hat{c}^{\dagger}_{i \uparrow} \hat{c}^{\dagger}_{i \downarrow} + \text{H.c.} \right)
\end{eqnarray}
where $\Delta$ is the superconducting gap.
The third term describes coupling of the magnetic impurity with the underlying lattice.
In our investigation, we describe the magnetic impurity captured by $H_\text{imp}$ as a classical spin aligned out-of-plane and only present on site [i.e. term proportional to $\delta ( r_{0} - r_{i} )$, where the subscript ``0'' denotes the impurity site].
In this case, the scattering potential at the position of the impurity is given as:
\begin{eqnarray}
H_\text{imp} =  K ( \hat{c}^{\dagger}_{0 \uparrow} \hat{c}_{0 \uparrow} + \hat{c}^{\dagger}_{0 \downarrow} \hat{c}_{0 \downarrow}) - J ( \hat{c}^{\dagger}_{0 \uparrow} \hat{c}_{0 \uparrow} - \hat{c}^{\dagger}_{0 \downarrow} \hat{c}_{0 \downarrow}) , 
\end{eqnarray}
where $K$ denotes the non-magnetic scattering potential, while $J$ denotes the coupling strength between the electrons and the magnetic impurity.
The classical magnetic impurity limit is technically achieved by taking $S \rightarrow \infty$ (large spin), while simultaneously letting $J \rightarrow 0$ so that $JS = \text{const}$~\cite{balatsky.vekhter.06}.
Effectively, the classical magnetic impurity acts on the system in two ways: (i) by shifting the chemical potential ($K$ term), that effectively leads to a modification of the number of electrons at site ``0''; (ii) by on-site Zeeman-like magnetic field ($J \equiv J S$ term, where $S$ is magnetic moment)~\cite{balatsky.vekhter.06}.

The Hamiltonian~(\ref{eq.ham}) describing the inhomogeneous problem, can be diagonalized via the following unitary transformation:
\begin{eqnarray}
\hat{c}_{i\sigma} = \sum_{n} \left( u_{in\sigma} \hat{\gamma}_{n\sigma} - \sigma v^{\ast}_{in\sigma} \hat{\gamma}_{n\bar{\sigma}}^{\dagger} \right)
\end{eqnarray}
where $\hat{\gamma}_{n}$ and $\hat{\gamma}^{\dagger}_{n}$ are quasi-particle fermionic operators, $u_{i n \sigma}$ and $v_{i n \sigma}$ are the eigenvector coefficients.
This leads to the Bogoliubov--de Gennes (BdG) equations~\cite{degennes.89}:
\begin{eqnarray}
\label{eq.2}
\mathcal{E}_{n\sigma} \left( \begin{array}{c}
u_{in\sigma} \\
v_{in\bar{\sigma}}
\end{array} \right) = \sum_{j} \left( \begin{array}{cc}
H_{ij\sigma} & D_{ij}  \\ 
D_{ij}^{\ast} & -H_{ij\bar{\sigma}}^{\ast}
\end{array} \right) \left( \begin{array}{c}
u_{in\sigma} \\
v_{in\bar{\sigma}}
\end{array} \right)
\end{eqnarray}
where, $H_{i j \sigma} = -t \, \delta_{\langle i,j \rangle} - [\mu+(K-\sigma \, J)\delta_{i 0}]\delta_{ij}$ and $D_{ij} = \Delta_{i} \delta_{ij}$, denotes the kinetic and superconducting part of the Hamiltonian, respectively~\cite{ptok.10,ptok.12,ptok.kapcia.15}.

From the solution of the BdG equations~(\ref{eq.2}) we can extract the local density of states (LDOS) for specified parameters of the system as~\cite{matsui.sato.03}:
\begin{eqnarray}
\nonumber \rho_{i} (\omega) = \sum_{n\sigma} \left[ | u_{in\sigma} |^{2} \delta \left( \omega - \mathcal{E}_{n\sigma} \right) + | v_{in\sigma} |^{2} \delta \left( \omega + \mathcal{E}_{n\bar{\sigma}} \right) \right] , \\
\end{eqnarray}
while the total DOS is given as $N (\omega) = \sum_{i} \rho_{i} ( \omega )$.
The LDOS for $\omega = \pm E_\text{YSR}$ denotes localization of the YSR state in real space~\cite{ptok.glodzik.17}, and can be useful in coherence length study.
In this case $\langle \rho_{i} (E_\text{YSR}) \rangle \propto \exp ( - r / \zeta_{c} )$ denotes the wavefunction of the YSR states and can be used to estimate the coherence length $\zeta_{c}$ for a given lattice and Fermi level.

Numerical computations have been done at zero temperature $T=0$ for the lattices with the periodic boundary conditions, containing $N_{a} \times N_{b} = 30 \times 30$ primitive unit cells.
In case of honeycomb lattice this corresponds to $1800$ sites, while for kagome and Lieb lattices, we have $2700$ sites.
For simplicity and without loss of generality, we assume constant value of $\Delta/t = 0.2$.
Additionally, to study only the role of DOS of the underlying system on the YSR states, we take $K = 0$.
In numerical determination, we have replaced the Dirac $\delta$ function by the Lorentzian $\delta(\omega) = \eta / [ \pi ( \omega^{2} + \eta^{2} ) ]$ with a small broadening $\eta = 0.05 t$.

\begin{figure*}
\centering
\includegraphics[width=\linewidth]{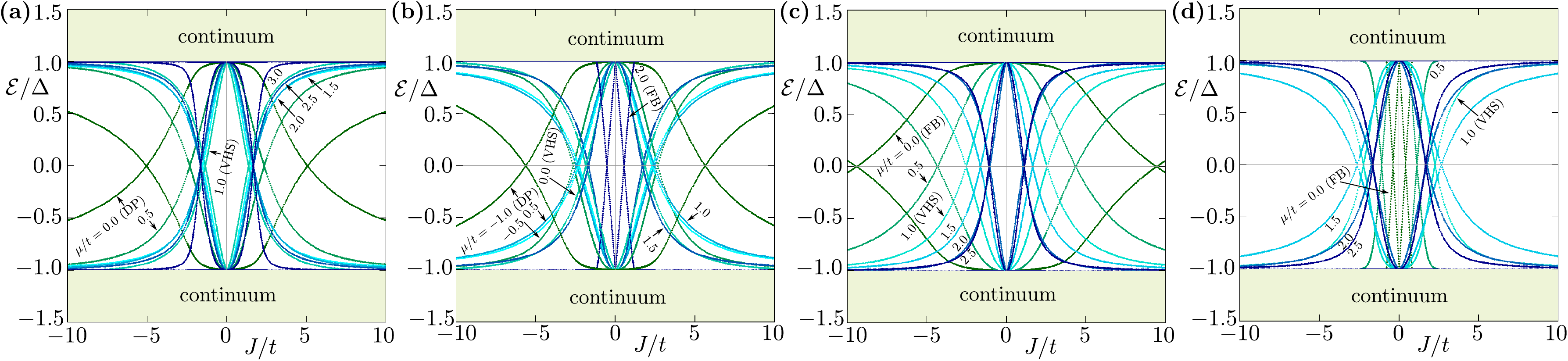}
\caption{
In-gap spectrum of the Shiba states for different values of the chemical potential (as labeled) in the case of: (a) hexagonal, (b) kagome, and (c,d) Lieb lattices.
For the Lieb lattice, the impurity was located in (c) the corner site or (d) the edge site [i.e. blue and green/red site in Fig.~\ref{fig.schemat}(c), respectively].
The chemical potentials for flat bands, van Hove singularities, and Dirac points, are marked as FB, VHS, and DP, respectively.
Additionally, green areas marked the ranges of energies for (out-gap) continuum states.
\label{fig.eysr_v_j}
}
\end{figure*}

\section{Numerical results and discussion}
\label{sec.num}

In our analysis, we consider honeycomb, kagome, and Lieb lattices.
Each of them is formed by the unit cells containing more than one site, and as a consequence in the DOS a few bands can be distinguished.
The honeycomb (kagome) lattice is formed by unit cells containing two (three) equivalent sites (in Fig.~\ref{fig.schemat} represented by dots with different colors).
In this case, one of the sublattices can be transformed to another by combination of a few translation, rotation, or reflection operations. 
Contrary to this, the Lieb lattice is formed by unit cells containing two different types of sites. 
Two edge sites [green/red dots in Fig.~\ref{fig.schemat}(c)] are equivalent to each other, but nonequivalent to the corner site [blue dots].
As a result, the Lieb lattice is characterized by two non-equivalent sublattices, the sublattice of edge site cannot be transformed to
the corner site sublattice, and {\it vice versa}.
We will show that this characteristic feature of the Lieb lattice plays an important role in realization of the Shiba states.

Depending on the lattice, in the DOS we can find a few interesting features.
For example, honeycomb and kagome lattices contain Dirac points (DP) at band touching points.
Similarly, in the DOS of the kagome and Lieb lattices the flat band (FB) feature can be distinguished.
Additionally, the van Hove singularities (VHS) in form of characteristic peaks in the DOS are visible for all lattices.
As we can see, the DOSs exhibit similar behaviors regardless of the chosen lattice. 
However, the DOS do not contain the full information about the lattice (e.g. symmetry of the system, number of neighboring sites, existence of eventual sublattices, etc.).
Despite the similarities in DOSs, the strong differences are well visible in the chemical potential dependence of the critical magnetic coupling $J_\text{c}$ (Fig.~\ref{fig.jc}) and the Shiba state energies $\pm E_\text{YSR}$ (Fig.~\ref{fig.eysr_v_j}).

First, we notice that the $J_\text{c}$ vs. $\mu$ plot follows the same symmetry as the DOS (cf.~Fig.~\ref{fig.jc} and Fig.~\ref{fig.schemat}).
For honeycomb and Lieb lattices, the DOS is symmetric with respect to the center of the band-width (related to $\mu/t= 0$) and the same character is reflected in $J_\text{c} (\mu)$.
Similarly, the asymmetric DOS of the kagome lattice is reflected in Fig.~\ref{fig.jc}(b).

$J_\text{c} ( \mu )$ exhibits a strong dependence on the position of the Fermi level which is related to $\mu$ (Fig.~\ref{fig.jc}).
However, dependence of $J_\text{c}$ on $N(E_\text{F})$ is highly unexpected.
In case of honeycomb and kagome lattices [presented in Fig.~\ref{fig.jc}(a) and Fig.~\ref{fig.jc}(b), respectively], the presence of the DP with $N(E_\text{F}) = 0$ leads to the occurrence of a peak in $J_\text{c}$.
Contrary to this, the presence of VHS leads to a relatively small reduction of $J_\text{c}$.
The most important modification of $J_\text{c}$ is introduced by the FB (theoretically with infinite DOS) in the kagome lattice.
In this case, $J_\text{c}$ decreases dramatically to its minimum value with respect to rest of the plot.
Nevertheless, most surprising results can be found in the Lieb lattice [Fig.~\ref{fig.jc}(c) and Fig.~\ref{fig.jc}(d)].
When impurity is located at the corner site,  $J_\text{c} ( \mu )$ exhibits features similar to the honeycomb lattice [cf.~Fig.~\ref{fig.jc}(a) and Fig.~\ref{fig.jc}(c)].
$J_{c}$ reaches the maximum value when $\mu$ is located at the FB.
Contrary to this, the impurity located at the edge site leads to opposite (expected) behaviors, i.e. dramatic decrease of $J_\text{c}$ for $\mu/t = 0$ (corresponding to the FB).

The results presented here are sensitive to the number of sites of the discussed system (cf. lines with different color in Fig.~\ref{fig.jc}).
Nevertheless, $J_\text{c}(\mu)$ has the same (qualitative) behavior independent of the size of the system, and describes the thermodynamic limit ($N_{a} \times N_{b} \rightarrow \infty$) relatively well for lattice with $30 \times 30$ unit cells.
Similar effects can be observed with decreasing $\Delta$, when emergence of the superconducting phase (i.e. when $\Delta$ is comparable to the gap between states) strongly modifies states in the system.

\begin{figure}[!b]
\centering
\includegraphics[width=\linewidth]{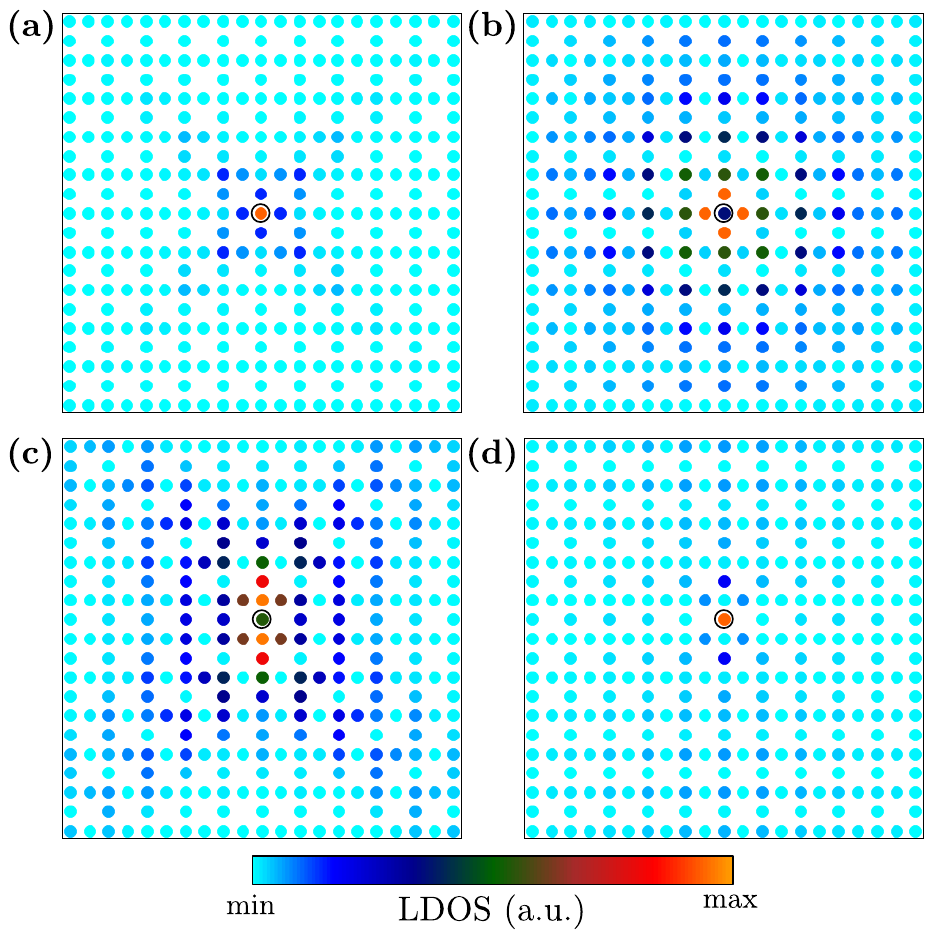}
\caption{
Local density of states (LDOS) for the Lieb lattice.
The position of the impurity is marked by a black circle. 
In panels (a) and (b) the impurity is located at the corner site [blue site in Fig.~\ref{fig.schemat}(d)], while in (c) and (d) it is located at the edge site [green or red site in Fig.~\ref{fig.schemat}(d)].
Results are shown for following values of parameters $(\mu/t;J/t)$: $(2.0;0.5)$, $(0.5;7.5)$, $(2.0;5.0)$, and $(0.5;0.5)$ for panel (a), (b), (c), and (d), respectively.
\label{fig.ldos}
}
\end{figure}

Value of $J_\text{c} (\mu)$ is related to the Shiba state energy $\pm E_\text{YSR}$ (Fig.~\ref{fig.eysr_v_j}).
Reduced value of $J_\text{c}$ induced by VHS or FB leads to a nearly linear dependence of $E_\text{YSR}$ on coupling strength $J$.
Similar features of the Shiba states in the presence of VHS were discussed by Uldemolins~{\it et~al.} in Ref.~\cite{uldemolins.mesaros.21}.
The Authors found strong reduction of $J_\text{c}$ when the Fermi level was located at the VHS. 
In this case, linearity of $E_{YSR}(J)$ was also reported around $J_\text{c}$.
In our case, the presence of a FB allows existence of the Shiba state with exactly linear $E_\text{YSR}(J)$ dependence [cf.~Fig.~\ref{fig.eysr_v_j}(b) or Fig.~\ref{fig.eysr_v_j}(d)].
For a system with small DOS (e.g. $\mu$ corresponding to the DP) or approximately constant value of DOS, the Shiba state energy has expected features, given by the general formula~(\ref{eq.eingap}).

Extraordinary properties of the Shiba states in the Lieb lattice can be directly connected  with the presence of two sublattices, formed by the corner sites or the edge sites.
The partial DOS projected on the corner or edge sites, which clearly show impact of each sublattices~\cite{slot.gardenier.17,jiang.huang.19,cui.zheng.20} could be a proof of this hypothesis.
In particular, whole spectral weight of the FB peak in the DOS corresponds to the edge site sublattice [giving vanishing PDOS for corner sites sublattice and opposite $J_\text{c}$ tuning for $\mu/t=0$, cf. Fig~\ref{fig.jc}(c) and Fig.~\ref{fig.jc}(d)].
Contrary to the FB peak, the DOS at the VHS has contribution from both sublattices~\cite{cui.zheng.20}.
This very well explains the $J_\text{c} (\mu)$ dependence for corner site sublattice [Fig.~\ref{fig.jc}(c)] and the edge site sublattice [Fig.~\ref{fig.jc}(d)].
Second direct proof can be given by the local DOS (LDOS) of the Shiba states induced by the magnetic impurities located at a given sublattice (Fig.~\ref{fig.ldos}).
As we can see, the pattern of the LDOS corresponding to the Shiba state strongly depends on the position of the impurity (cf. top and bottom panels on Fig.~\ref{fig.ldos}).
Similarly, the coherence length depends on the parameters of the systems (cf. left and right panels on Fig.~\ref{fig.ldos}).
From these numerical studies we can find that the Shiba state is mostly localized at the same sublattice as the magnetic impurity site (marked by black circle). 
Only in the sites, which are neighbors to the impurity's location, some modification of LDOS is observed.
Interestingly, similar behavior was observed experimentally within the STM measurements of the Zn impurity in copper--dioxide Lieb lattice of the high temperature superconductor Bi$_{2}$Sr$_{2}$CaCu$_{2}$O$_{8+\delta}$~\cite{pan.hudson.00}.
In this case, for the impurity located at the corner site, the LDOS of bound states was observed in diagonal directions (rotated $45^{\circ}$ to $x$-$y$ axis), i.e. situation similar to Fig.~\ref{fig.ldos}(a).

\begin{figure}[!t]
\centering
\includegraphics[width=\linewidth]{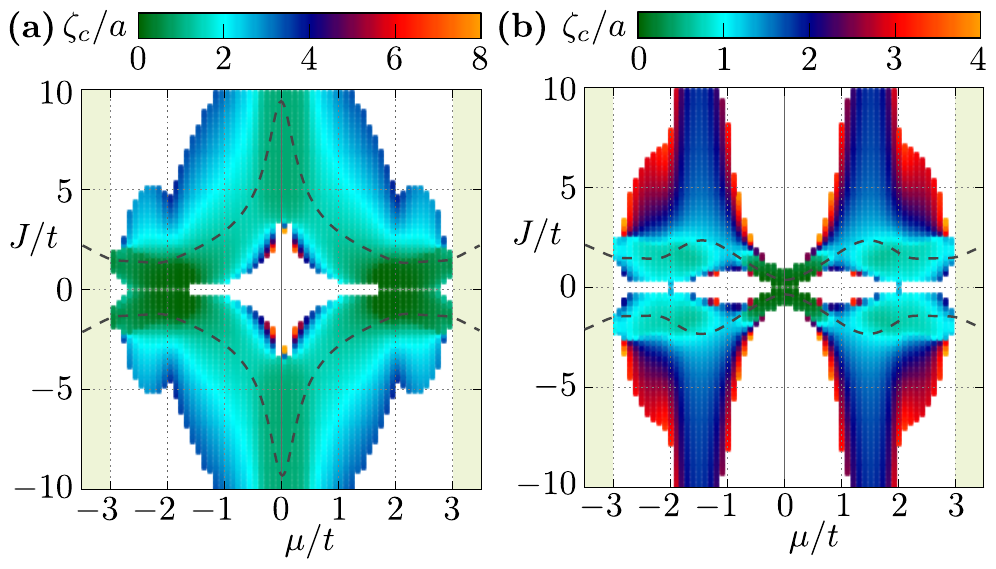}
\caption{
Coherence length $\zeta_{c}$ as a function of chemical potential $\mu$ and magnetic coupling $J$ for the Lieb lattice when the impurity is located at (a) the corner site or (b) the edge site [i.e. blue and green/red site in Fig.~\ref{fig.schemat}(c), respectively].
Dashed black line denotes the value of critical $J$ for a given $\mu$ (cf.~Fig.~\ref{fig.jc}).
Results are presented for Shiba states with energies $|E_\text{YSR}| < 0.95\Delta$.
$a$ is the distance between neighboring sites (i.e. distance between corner and edge state), taken as a unit of distance.
\label{fig.zeta}
}
\end{figure}

Exponential decay of the Shiba states as a  function of distance from the impurity gives information about the coherence length $\zeta_\text{c}$ (Fig.~\ref{fig.zeta}).
Similar to $J_{c}$ for the Lieb lattice, the coherence length depends on the sublattice in which the magnetic impurity is located.
Moreover, around $J_{c}(\mu)$ (represented by black dashed line) the coherence length is relatively small (in range $\sim 2a$ or $\sim 1a$ for impurity in the corner site or edge site, respectively).
For $J \gg J_{c}$ the coherence length can be much bigger -- this behavior is also visible in LDOS discussed earlier (Fig.~\ref{fig.ldos}).


\section{Summary}
\label{sec.sum}

In this paper we discuss the effect of density of states singularities on the Shiba states.
In particular, we investigated the role of Dirac point, van Hove singularity, and flat band in honeycomb, kagome, and Lieb lattices.
In its simplest form, the energy of the Shiba states strongly depends on the density of states at the Fermi level.
For example, presence of a flat band leads to strong suppression of the critical magnetic coupling $J_\text{c}$, while existence of the Dirac point leads to an enhanced  $J_\text{c}$.
We examined the tuning of the parameters (e.g. energy or critical magnetic coupling) describing the Shiba states in aforementioned lattices.

The Shiba states realized in the honeycomb and the kagome lattices exhibit typical behaviors.
For the Fermi level at the Dirac point we observed maximum of $J_\text{c}$, while the flat band strongly suppresses $J_\text{c}$.
Similarly, the van Hove singularity leads to a decrease in  $J_\text{c}$.
Contrary to this, the Lieb lattice containing two sublattices (of the corner and edge sites), exhibit extraordinary behavior.
In this case, the properties of the Shiba states strongly depend on the position of the magnetic impurity (in a specific sublattice).
The atypical behavior is observed in $J_\text{c}$, the local density of states, and coherence length studies.
{\it (i)} The magnetic coupling exhibits dependence similar to the density of states projected on the specific sublattice.
{\it (ii)} Majority of the spectral weight in local density of states of the Shiba states is observed in the sublattice containing the magnetic impurity.
{\it (iii)} The coherence length strongly depends on the sublattice in which the magnetic impurity is located.
These findings should be generic for systems with a few sublattices, and can explain some experimental data observed in systems where Lieb lattice is realized (like high temperature superconductors).


\begin{acknowledgments}
We kindly thank Szczepan G\l{}odzik, Przemys\l{}aw Piekarz, and Pascal Simon for insightful discussions.
This work was supported by National Science Centre (NCN, Poland) under Projects No.
2017/25/B/ST3/02586 (S.B.), 
and
2017/24/C/ST3/00276 (A.P.). 
In addition, A.P. appreciates funding in the frame of scholarships of the Minister of Science and Higher Education (Poland) for outstanding young scientists (2019 edition, No.~818/STYP/14/2019).
\end{acknowledgments}

\bibliography{biblio}

\end{document}